\documentclass[useAMS,usenatbib,usegraphicx]{mn2e}

\usepackage{amssymb}

\newcommand{\var}{\mathop{\rm Var}\nolimits}
\newcommand{\cov}{\mathop{\rm Cov}\nolimits}
\newcommand{\trace}{\mathop{\rm Tr}\nolimits}

\newcommand{\expect}{\mathbb{E}}

\newcommand{\const}{\mathop{\rm const}\nolimits}
\newcommand{\FAP}{{\rm FAP}}
\newcommand{\As}{{\rm As}}
\newcommand{\Ex}{{\rm Ex}}

\title[Accounting for velocity jitter in planet search surveys]%
{Accounting for velocity jitter in planet search surveys}
\author[R.V.~Baluev]%
{Roman V.~Baluev \thanks{E-mail: roman@astro.spbu.ru}\\
Sobolev Astronomical Institute, St Petersburg State University,
Universitetskij prospekt 28, Petrodvorets, St Petersburg 198504, Russia}
\begin{document}

\date{  Accepted 2008 November 6.
        Received 2008 November 5;
        in original form 2007 December 22}

\pagerange{\pageref{firstpage}--\pageref{lastpage}} \pubyear{2008}

\maketitle

\label{firstpage}

\begin{abstract}
The role of radial velocity (RV) jitter in extrasolar planet search surveys is discussed.
Based on the maximum likelihood principle, improved statistical algorithms for RV fitting
and period search are developed. These algorithms incorporate a built-in jitter
determination, so that resulting estimations of planetary parameters account for this
jitter automatically. This approach is applied to RV data for several extrasolar planetary
systems. It is shown that many RV planet search surveys suffer from periodic systematic
errors which increase effective RV jitter and can lead to erroneous conclusions. For
instance, the planet candidate HD74156~d may be a false detection made due to annual
systematic errors.
\end{abstract}


\section{Introduction}
\label{sec_intro}
When analysing radial velocity (RV) data from planet search surveys, we should bear in
mind that total uncertainties of these RV measurements are assembled from instrumental
uncertainties and a `jitter'. Partly, this jitter is produced by various processes on the
star leading to instabilities of the observed radial velocity. Estimations of planetary
masses and orbital elements depend on full RV uncertainties, hence RV jitter should be
accounted for in the data analysis. Usually, empirical models based on a set of stellar
characteristics are used to assess RV jitter \citep[e.g.,][]{Wright05,Saar98}.
Unfortunately, this way of jitter estimation allows accuracies of $\sim 50\%$ only or even
worse. Often, the RV jitter remains almost unconstrained a priori (in comparison with
instrumental errors) and represents an extra unknown parameter.

It is worth stressing that the jitter also depends on the instrument, on the way of
observations and obtaining final RV measurements. For instance, sufficiently long
exposures average out stellar oscillations and decrease the apparent RV jitter. Extra
systematic errors (which have not yet been investigated in detail in planet search
surveys) should increase it. When performing a joint analysis of data from different
observatories, we must not forget that their effective RV jitter may be quite different,
implying different statistical weights to their RV data.

Until accurate a priori estimations of RV jitter are constructed, we need to use some
statistical algorithm of data analysis, which could account properly for the presence of
poorly known RV jitter. It is possible to construct a statistical jitter estimation based
on the scattering of the data around the RV model for a given star. The aim of this paper
is to propose efficient tools implementing this idea. The algorithm can be organised so
that the jitter estimation is automatically accounted for in estimations of planetary
masses and orbital parameters (and vice versa).

In Section~\ref{sec_jitter}, the background connected with RV jitter is outlined. In
Sections~\ref{sec_ls} and~\ref{sec_mom}, the traditionally used algorithms of RV curve
fitting and posterior empirical jitter determination are briefly discussed and are shown
to be unsuitable for our goals. In Section~\ref{sec_maxlik}, the maximum likelihood
approach is proposed for joint estimation of RV jitter and parameters of the RV curve. It
is shown that this approach can take into account the presence of unknown RV jitter
properly. In Section~\ref{sec_bias}, a modification of the likelihood function is
introduced. This modification allows to perform a `preventive' reduction of the
statistical bias in the RV jitter. Several other issues connected with biasing of
estimations are also discussed in this section. In Section~\ref{sec_nongauss}, the effect
of possible non-Gaussian distribution of RV errors is considered. It is shown that many
important properties of the maximum likelihood algorithm constructed in the paper, are not
destroyed by non-Gaussian nature of RV errors. In Section~\ref{sec_numcalc}, an efficient
numerical implementation of the analytic algorithm is described. This implementation is
based on the common non-linear Levenberg-Marquardt-Gauss least squares algorithm. In
Section~\ref{sec_test}, the modified likelihood ratio test is proposed for checking
consistency of RV models emerging in planet searches with RV data. This test incorporates
a built-in estimation of the RV jitter. Based on this test, a generalisation of the
Lomb-Scargle periodogram is proposed in Section~\ref{sec_periods}. Results obtained for
several RV datasets from current planet search surveys are presented in
Section~\ref{sec_apps}.

\section{Radial velocity jitter}
\label{sec_jitter}
High-precision RV data from planet search surveys suffer from the phenomenon called `RV
jitter'. RV measurements often shows scattering far beyond the level which is expected
from their internal uncertainties. Let $v_i$ denote $N$ RV measurements made at the epochs
$t_i$. Denoting the internal standard errors of $v_i$ as $\sigma_{{\rm meas}, i}$, the
total variances of RV error are usually derived as
\begin{equation}
 \sigma_i^2 = \sigma_{{\rm meas}, i}^2 + \sigma_\star^2,
\label{SquareModel}
\end{equation}
where the constant term $\sigma_\star^2$, softening differences between $\sigma_i$,
characterizes the RV jitter.

It is necessary to clarify the notion `jitter'. We will name the term $\sigma_\star^2$
in~(\ref{SquareModel}) as `jitter' (or `RV jitter', `full RV jitter') regardless its
physical nature. In the astrophysical part, the RV jitter is inspired by various processes
in the star leading to an apparent instability of its radial velocity. Also, it depends on
the instrument, on the way of observation and its reduction to final radial velocity
measurement. For example, an exposure as long as $20-30$~min averages out apparent RV
variations inspired by stellar oscillations, which have periods of several minutes for
solar-like stars \citep{Mayor03,OToole08}. This decreases the astrophysical part of the
full RV jitter. However, the astrophysical jitter does not represent the only source of RV
variations beyond the expected noise level. Other sources like extra systematic RV errors
lie in the instrumentation and in the data reduction (but they may depend on stellar
properties as well). In Section~\ref{sec_apps} we will see that effective RV jitter may be
quite different for different observatories. Note that imperfection of RV models (say,
extra Doppler variability due to undetected planets in the system) also increase the full
jitter, but this increase does not depend on an instrument.

It is worth stressing that we are not intending to find here any temporal RV model for the
jitter. We model the RV jitter in the statistical sense, using the square-additive model
of RV uncertainties. This simplification should yield reliable results for the case, when
the jitter has roughly uniform frequency spectrum in the frequency range that we are
interested in. In planet searches, we are mostly interested in periods of RV variations
from days to years. This means that, for example, the stellar oscillations investigated by
\citet{Mayor03} and \citet{OToole08} quite can be processed in this way, because the range
of their periods (minutes or even hours) lies far beyond the period range that we deal
with. However, some kinds of extra RV variability in the data may require an explicit
representation in the temporal model of the RV curve. These include, for example,
quasi-periodic long-period instrumental errors (see Section~\ref{sec_apps}), RV drifts
inspired by spots on the rotating stellar surface
\citep{Bonfils07,SaarDonahue97}.

\section{Least squares approach}
\label{sec_ls}
If we knew the exact statistical weights of the observations, $w_i \propto \sigma_i^{-2}$,
we could write down the full variances of $v_i$ as
\begin{equation}
 \sigma_i^2 = \kappa/w_i,
\label{MultModel}
\end{equation}
where the parameter $\kappa$ (the error variance for the unit weight) is unspecified. This
is the framework which is typically assumed in usual statistical algorithms. Clearly, the
models~(\ref{MultModel}) and~(\ref{SquareModel}) are different. Hereafter, we will refer
to~(\ref{SquareModel}) as to the `square-additive' model and to~(\ref{MultModel}) as to
the `multiplicative' one. Both these models require a sequence of $N$ a priori fixed
quantities ($\sigma_{{\rm meas}, i}$ or $w_i$) and contain an unknown `variance' parameter
($\sigma_\star$ or $\kappa$). If all instrumental uncertainties are equal to each other
then the models~(\ref{SquareModel}) and~(\ref{MultModel}) become equivalent.

We need to fit our RV observations by a model $v=\mu(t,\btheta)$ which depends on $d$ free
parameters forming the vector $\btheta$. Traditionally, the best-fitting estimations
$\btheta^*$ are obtained in result of minimizing the function $\chi^2 = \left\langle (v -
\mu)^2/\sigma^2 \right\rangle$ by $\btheta$.\footnote{See Appendix~\ref{sec_not} for
explanations of several mathematical notations (like the operation $\langle * \rangle$)
used in the paper.} This is equivalent to minimizing the function $\tilde\chi^2 = \kappa
\chi^2 = \left\langle w (v-\mu)^2 \right\rangle$ which does not contain any undefined
quantities. This is the essence of the least squares principle commonly used to obtain the
best-fitting values of unknown parameters of the RV curve.

The least squares approach assumes that the weights of observations and, hence, the RV
jitter are known a priori. This a priori jitter estimation is usually obtained from
empirical models \citep{Saar98,Wright05} or even is neglected. Inaccurate values of the
jitter inject extra bias in the least squares estimations and decrease their reliability,
especially for the cases when the planetary orbits are not constrained well. Still, the
accuracy of the a priori jitter estimations is not better than $\sim 50\%$. The jitter of
several m/s (that is, of the order of typical internal RV precision reached in planet
search surveys) have the largest effect on the best-fitting parameters of the RV curve.
Unfortunately, it is the region where the a priori RV jitter estimations are mostly
uncertain.

\section{Method-of-moments estimator}
\label{sec_mom}
If the true values of the parameters $\btheta$ of the RV curve were somehow known, we
could estimate RV jitter based on the observed scattering of the residuals around the RV
model $\mu(t,\btheta)$ as follows:
\begin{equation}
 \sigma_\star^2 = \left\langle (v-\mu(t,\btheta))^2 \right\rangle/N -
 \left\langle \sigma_{\rm meas}^2 \right\rangle/N.
\label{EstMom}
\end{equation}
It easy to see that the first term in the right hand side of this equation represents the
second sample moment of the residuals. Its mathematical expectation is $\sigma_\star^2 +
\langle\sigma_{\rm meas}^2\rangle/N$. Therefore, the estimator~(\ref{EstMom}) could be
obtained after equating the second sample moment to its expectation. Such an estimator is
called the method-of-moments estimator (MME). The jitter estimation~(\ref{EstMom}) is,
probably, the most easy and intuitive one. However, it does not estimate anything but the
RV jitter. Eventually, we are interested in the estimation of $\btheta$, taking into
account some most suitable value of the RV jitter. It is possible to organise an iterative
process based on the MME of the RV jitter and on the least squares estimator of $\btheta$,
but this way requires intensive calculations due to multiple non-linear $\chi^2$
minimizations and thus is not practical. In addition, estimators constructed using the
method of moments do not necessary provide the best accuracy. It is not hard to show that
the variance of the MME~(\ref{EstMom}) is $2\langle\sigma^4\rangle/N^2$. As will be shown
in Section~\ref{sec_maxlik}, this is not the minimum variance possible for estimating the
RV jitter.

\section{Maximum likelihood estimator}
\label{sec_maxlik}
We note that the least squares principle is often considered as a special case of a more
general maximum likelihood principle. Assuming that the errors of the RV measurements are
uncorrelated and Gaussian, we can write down the associated log-likelihood function as
\begin{equation}
 \ln \mathcal L = - \chi^2/2 - \langle \ln\sigma \rangle + N \ln\sqrt{2\pi}.
\label{log-likelihood}
\end{equation}
This function depends on the parameters $\btheta$ and $\kappa$. The maximum likelihood
principle implies that estimations of these parameters correspond to the maximum value of
$\mathcal L$ (or, equivalently, $\ln\mathcal L$). The function~(\ref{log-likelihood}) can
be rewritten in the form $\ln \mathcal L = - \tilde\chi^2/(2\kappa) - (N \ln\kappa)/2 -
\langle \ln w \rangle / 2 + \const$. When the weights $w_i$ are fixed and known, the
maximization of $\ln \mathcal L$ can be performed elementary. The resulting value of
$\btheta^*$ is given by the least squares estimator. For the estimation of $\kappa$, we
obtain the well-known result $\kappa^* = \tilde\chi^2(\btheta^*)/N$.

Let us now assume that we have $r$ `variance' parameters $\bmath p$ (say, RV jitter of a
given star observed with different instruments) entering in the model of $\sigma_i$. Let
us denote the full vector of $d+r$ unknown (or at least poorly known) parameters
$(\btheta,\bmath p)$ as $\bxi$. For the square-additive model of $\sigma_i$, we should
maximize $\ln\mathcal L$ by $\btheta$ and $\bmath p$ simultaneously. The values
$\btheta^*$ and $\bmath p^*$ providing the maximum of $\ln \mathcal L$, represent the
joint maximum likelihood estimator (MLE) $\bxi^*$. It is important that information about
$\bmath p^*$ is automatically accounted for in the estimation $\btheta^*$, and vice versa.
An analytic maximization of $\ln\mathcal L$ for the model~(\ref{SquareModel}) does not
seem possible. However, an effective way of numerical maximization of the likelihood
function will be described in Section~\ref{sec_numcalc}.

Any estimation is not of much use without associated uncertainty, i.e. without estimation
of its variance. The variance-covariance matrix of an MLE is usually expressed using the
Fisher's information matrix
\begin{equation}
\mathbfss F(\bxi) = \expect\left( \frac{\partial\ln\mathcal L}{\partial \bxi} \otimes
               \frac{\partial\ln\mathcal L}{\partial \bxi} \right) =
               - \expect\left( \frac{\partial^2\ln\mathcal L}{\partial\bxi^2} \right),
\end{equation}
calculated for the true value of $\bxi$. The inverse $\mathbfss F^{-1}$ represents an
asymptotic ($N \to \infty$, i.e. large sample) approximation to $\var\bxi^*$
\citep[][\S~6.4]{Lehman-est}. In our case, the Fisher's information matrix can be written
in the block form
\begin{equation}
\mathbfss F = \left(\begin{array}{cc}
\mathbfss F_{\btheta\btheta} & \mathbfss F_{\bmath p\btheta}\\
\mathbfss F_{\btheta\bmath p} & \mathbfss F_{\bmath p\bmath p}\\
\end{array}\right),
\label{LikHess}
\end{equation}
where the sizes of the submatrices match the dimensions of the vectors marked in
subscripts. The calculation of $\mathbfss F$ yields, in particular, that $\mathbfss
F_{\btheta\bmath p} = 0$ and that $\mathbfss F_{\btheta\btheta}$ coincides with the
Fisher's information matrix for the least squares estimator:
\begin{equation}
\mathbfss Q = \left\langle \mu'_\btheta \otimes \mu'_\btheta/\sigma^2 \right\rangle
\label{matrQ}
\end{equation}
This implies that the vectors $\btheta^*$ and $\bmath p^*$ are asymptotically uncorrelated
and the asymptotic variance-covariance matrix of $\btheta^*$ is the same as in the usual
least squares approach. If our dataset is merged from several time series obtained at
different observatories, we may be interested in separate estimations of RV jitter. It is
not hard to show that these separate jitter estimations are asymptotically uncorrelated
also. Finally,
\begin{eqnarray}
\var \btheta^* \simeq \mathbfss Q^{-1},\qquad
\var p_j^* \simeq \varepsilon_j^2 = 2/\langle\sigma^{-4}\rangle_j,
\label{Var}
\end{eqnarray}
where the index $j$ means that the respective summation $\langle*\rangle$ should be
restricted to the $j^{\rm th}$ sub-dataset. The only seeming obstacle in practical use
of~(\ref{Var}) comes from the fact that formally we should substitute the true values of
parameters $\btheta$ and $\bmath p$ in these equations. In practice, we can substitute
only the estimations $\btheta^*,\bmath p^*$, which we have obtained before. This is
admissible for calculating the asymptotic large sample approximation of $\var\bxi^*$,
because the estimations tend to the true values when $N$ grows.

The MLEs possess many good statistical properties when the number of observations is
large. Under certain regularity conditions, they are asymptotically ($N\to\infty$)
unbiased (but see Section~\ref{sec_bias} for some cautions), asymptotically Gaussian and
asymptotically efficient \citep[][chapter~6]{Lehman-est}.\footnote{This behaviour can be
damaged in an incarefully chosen parametrization. For instance, it is a frequent case for
hot Jupiter planets when the orbital eccentricity estimation looks like $e=0.05\pm 0.05$
and the argument of the periastron $\omega$ is ill-determined. Then the distribution of
$e$ and $\omega$ is non-Gaussian. This is due to the formal singularity of the point $e=0$
in the polar coordinate system $(e,\omega)$. This trouble is easy to overcome by means of
the change of variables $x=e\cos\omega, y=e\sin\omega$. The joint distribution of $(x,y)$
is already close to the bivariate Gaussian one.} The latter property means that the
statistical uncertainties of MLEs approach the minimum possible ones when $N$ grows.
Comparing the uncertainty of the MLE, $p_j$, with the uncertainty of the MME from
Section~\ref{sec_mom}, we can obtain that their ratio is equal to $\sqrt{\langle \sigma^4
\rangle \langle \sigma^{-4} \rangle}/N$. Due to the Cauchy-Schwarz inequality, this
quantity is not less than $1$. This means that the MLE yields generally more accurate
estimation of the RV jitter than the MME. For instance, the RV uncertainties of the Lick
data for 51 Pegasi (see Section~\ref{sec_apps}) imply roughly double advantage of the MLE.

The MLE is organised so that the resulting value of $\chi^2/N$ is always close to unity.
This means that the $\chi^2$ statistic can no longer be used as a measure of the fit
quality. Instead, we should use some other statistic, based on the full likelihood
function~(\ref{log-likelihood}). For this statistic to be intuitively clear, it should be
measured in the same units as $v_i$. Therefore, it should be proportional to $\mathcal
L^{-1/N}$. To find a suitable proportionality factor, let us assume for a moment that
$\chi^2=N$ exactly. Then $\mathcal L^{-1/N} = \sigma_{\rm geom} e^{0.5}\sqrt{2\pi}$, where
$\sigma_{\rm geom}$ is the geometric mean of $\sigma_i$. Therefore, already for the
general case, we may introduce the following likelihood goodness-of-fit statistic:
\begin{equation}
l = \mathcal L^{-1/N} e^{-0.5}/\sqrt{2\pi} \approx 0.2420 \mathcal L^{-1/N}.
\label{LikGOF}
\end{equation}
This statistic describes naturally the overall scattering of RV measurements around a
given RV model, for a given value of the RV jitter.

\section{Bias reduction}
\label{sec_bias}
It is well-known that linear least squares estimations are `unbiased', i.e. their
mathematical expectations are equal to true values of parameters. This property is very
important, because it allows us to hope that such estimations are related to true values
at all. Both square-additive and multiplicative models of RV uncertainties require
non-linear likelihood maximization to estimate the noise level parameter ($\kappa$ of
$\sigma_\star^2$). In general, maximum likelihood estimations are biased, but their bias
tends to zero as $N \to \infty$ \citep[\S~6.4]{Lehman-est}. Nevertheless, the biasing for
real RV time series with finite $N$ may become practically significant and may require a
reduction. For instance, it is well-known that the maximum likelihood estimation $\kappa^*
= \tilde \chi^2(\btheta^*)/N$, derived in Section~\ref{sec_maxlik}, is biased by $\mathcal
O(1/N)$ and the unbiased estimation is $\kappa^* = \tilde \chi^2(\btheta^*)/ (N-d)$. In
practice, we may quite have a set of $d\sim 20$ parameters of the Keplerian RV curve (for
a four-planet system) with $N\sim 100$ observations only. In this case, the relative bias
in $\kappa^*$ (about $d/N \sim 20\%$) exceeds the relative uncertainty of $\kappa^*$
(about $1/\sqrt N \sim 10\%$). We may expect a similar biasing for $\sigma_\star^2$. The
general reason of this biasing comes from the fact that residuals underestimate true
errors in average. This underestimation increases when the number of free parameters
grows. As an illustration, in the extremal case $d=N$ we could plot a model curve
transiting through all the data points exactly. In this case, all residuals would vanish.

We need to reduce jitter bias so that the resulting estimation of $\btheta$ would account
for this reduction automatically. This reduction can be reached by means of proper
$\mathcal O(1/N)$ modification of the functions~(\ref{log-likelihood}) and~(\ref{LikGOF}).
This is the approach of `preventive' bias reduction \citep{Firth93}. For our specific
goal, the likelihood function should be modified so that the residuals should be increased
by the relative quantity $\sim d/N$, in order to reach a more accurate representation of
measurement errors. The following modification looks convenient in practice:
\begin{eqnarray}
 \ln \tilde{\mathcal L} = - \chi^2/(2\gamma) - \left\langle \ln\sigma \right\rangle + N \ln\sqrt{2\pi},
\label{log-likelihood-mod} \\
 \tilde l = \tilde{\mathcal L}^{-1/N} e^{-0.5}/\sqrt{2\pi} \approx 0.2420 \tilde{\mathcal L}^{-1/N},
\label{LikGOF-mod}
\end{eqnarray}
where $\gamma = 1 - d/N$. Clearly, such $\mathcal O(1/N)$ modification should not destroy
the large-sample properties (like asymptotic normality and asymptotic efficiency) of the
maximum likelihood estimator. But moderate- and small-sample properties look now better.
Maximizing~(\ref{log-likelihood-mod}) instead of~(\ref{log-likelihood}) kills all bias in
the estimation of $\kappa$ for the multiplicative model of RV uncertainties. In the case
of the square-additive model, some residual bias may remain. This remaining bias can be
calculated till the first order, $\mathcal O(1/N)$, analytically, using cubic part of the
Taylor expansion of $\ln\mathcal L(\bxi)$ near the true value of $\bxi$ \citep[see,
e.g.,][]{CoxSnell68,Firth93}. These calculations involve quite bulky tensor algebra and
are omitted here. The final result (for the case $r=1$) looks like
\begin{equation}
 ({\rm correction~to}~\sigma_\star^2) = \frac{1}{\nu} \left(\trace\left(\mathbfss Q^{-1}
   \widetilde{\mathbfss Q} \right) - \lambda \frac{d}{N} \right),
\label{jitter-bias}
\end{equation}
where
\begin{equation}
 \widetilde{\mathbfss Q} = \left\langle \mu'_\btheta\otimes \mu'_\btheta/\sigma^4 \right\rangle,
 \qquad \lambda = \left\langle \sigma^{-2} \right\rangle,
 \quad \nu = \left\langle \sigma^{-4} \right\rangle.
\label{matrQP}
\end{equation}
The counterbalancing term in the equality~(\ref{jitter-bias}), containing $d/N$, was
produced by our modification of the likelihood function. If all $\sigma_i$ are equal to
each other then the correction~(\ref{jitter-bias}) is zero, as we could expect (recall
that the multiplicative model of $\sigma_i$ is equivalent to the square-additive one for
this case). When RV jitter is estimated separately for different components of the
combined time series, we should apply this bias correction separately as well. In this
case, it is necessary to restrict summations $\langle * \rangle$ in~(\ref{matrQP}) over
the respective sub-datasets, but to keep the full summation for the matrix $\mathbfss Q$.
The built-in correction provided by the likelihood function
modification~(\ref{log-likelihood-mod}) normally accounts for a large fraction of the bias
in jitter. Therefore, the cross influence of this bias on the estimations of $\btheta$ is
significantly decreased.

To correct the bias in estimations, the algorithm proposed by \cite{Quenouille56} may be
used. This is also called the `Jackknife' or `leave-one-out' method. It is as follows:
\begin{enumerate}
\item Calculate the basic (biased by $\mathcal O(1/N)$) estimation $x$ of a given
parameter $\xi$ from the full set of $N$ observations.

\item Construct $N$ reduced time series with $i^{\rm th} (i=1,2,\ldots N)$ measurement
omitted. Therefore, each reduced time series should consist of $N-1$ data points.

\item Calculate $N$ new estimations $x'_i (i=1,2,\ldots N)$ of $\xi$ by re-fitting
with every of the reduced time series. The bias of $x'_i$ will be about $\sim 1/(N-1)$,
hence these new estimations will be shifted with respect to $x$ by about $\sim
(1/(N-1)-1/N) = \mathcal O(1/N^2)$.

\item Calculate the sum $b_1 = \sum_{i=1}^N (x'_i-x)$. The result $b_1 = \mathcal
O(1/N)$ is the first-order bias of $x$. That is, the corrected estimation $x-b_1$ should
be biased by $\mathcal O(1/N^2)$ only.
\end{enumerate}
The main advantage of this algorithm is that its implementation is model-independent and
easy. Also, this algorithm does not require for the distribution of RV errors to be
Gaussian. In addition, it can be directly applied to either `variance' ($\bmath p$) or
usual ($\btheta$) parameters. Unfortunately, it is rather time-consuming because it
requires many non-linear fits.

\section{Non-Gaussian errors}
\label{sec_nongauss}
To write down the equality~(\ref{log-likelihood}) for the likelihood function, we have
assumed that RV errors follow Gaussian distributions. Some fears are sometimes expressed
that RV errors in planet search surveys may be significantly non-Gaussian
\citep[e.g.,][]{Marcy05,Butler06}. Then, strictly speaking, the
function~(\ref{log-likelihood}) is not a likelihood function and estimations obtained from
its maximization may be shifted with respect to the true MLE. Usually we have not enough
information to construct the true likelihood function. Then the usage of simple Gaussian
likelihood functions like~(\ref{log-likelihood}) or (\ref{log-likelihood-mod}) may be
reasonable. This is called sometimes the `pseudo maximum likelihood' approach
\citep[][\S~4.18]{Bard}.

How much the non-gaussianity of RV errors can affect the properties of the estimations
obtained using the Gaussian likelihood function~(\ref{log-likelihood}) and its
modification~(\ref{log-likelihood-mod})? To get some preliminary answer to this question,
let us consider a simplified situation of the least-squares algorithm from
Section~\ref{sec_ls} with RV model being linear with respect to unknown parameters. The
class of linear models incorporate, for instance, sinusoidal signals ($C\cos\omega t +
S\sin\omega t$ with \emph{a priori} fixed frequency $\omega$ but free linear parameters
$C$ and $S$), and polynomial trends. This is the well-known linear regression problem. The
associated linear least-squares estimations can be expressed explicitly as certain linear
combinations (or weighted sums) of the observations, regardless the shape of the input
errors distribution. The general expressions for the coefficients are too unpleasant to be
written down here, but they can be easily found in any textbook on the least-squares
method. The errors of the derived linear estimations represent just the same linear
combination of the observational errors (again regardless the degree of their
gaussianity). This immediately implies the following properties of the linear
least-squares estimations in the non-Gaussian situation:
\begin{enumerate}
\item If our RV model is correct, such estimations are \emph{exactly} unbiased, regardless
the shape of the distribution of the input RV errors.

\item If the variances of the input RV errors exist (they may not exist, e.g., for
heavy-tail Cauchy distribution) and are correctly modelled, the variances and correlations
of derived estimations are \emph{exactly} the same as in the case of Gaussian errors. In
the non-Gaussian case, the linear least-squares estimator is no longer guaranteed to be
strictly efficient, but still its variance is minimum possible among all unbiased
\emph{linear} estimators (the Gauss-Markov theorem).

\item If the conditions of the central limit theorem for the given distribution of RV
errors are satisfied, the joint distribution of the derived estimations tends to the
multivariate Gaussian one when $N\to\infty$.
\end{enumerate}
The mentioned general properties of the least-squares estimators are well-known in
statistics \citep[e.g.,][\S 23.2.6]{Koroluk}.

Of course, the models of the RV curve met in planet search syrveys typically incorporate
non-linear Keplerian RV functions. We should not expect that the nice properties of the
linear least-squares estimations with non-Gaussian input errors should hold true for the
more complicated non-linear pseudo maximum likelihood case. However, we can suspect that
at least some of these properties may be conserved approximately in the asymptotic sense
for $N\to\infty$. This problem was considered rigorously by \citet{Gourieroux84} (pay
particular attention to their Section~6). One may be surprised, that (of course under
certain regularity conditions) many important asymptotic properties of maximum likelihood
estimators, constructed for Gaussian errors, are conserved in the pseudo maximum
likelihood case, i.e. when the errors do not follow Gaussian distributions. For example,
the pseudo maximum likelihood estimators are asymptotically unbiased and Gaussian.
However, the asymptotic efficiency may be lost: we cannot construct even asymptotically
efficient estimator if the shape of the distributions of the RV errors is not known
precisely. The asymptotic variance-covariance matrix of estimations $\btheta^*, \bmath
p^*$ in the case of non-Gaussian errors can be derived from the formulae given in the
Appendix~5 of the paper by \citet{Gourieroux84}. The matrix $\var\btheta^* \simeq
\mathbfss Q^{-1}$ is unchanged (in the asymptotic large-sample approximation). Jitter
estimations corresponding to different observatories are uncorrelated again. However, a
non-zero skewness of RV errors inspires some correlation between $\btheta^*$ and $p_j^*$,
and an excess kurtosis distorts the variances of $p_j^*$:
\begin{eqnarray}
 \cov(\btheta^*,p_j^*) &\simeq& \mathbfss Q^{-1} \left\langle \As \frac{\mu'_\btheta}{\sigma^3}
       \right\rangle_j \frac{\varepsilon_j^2}{2}, \nonumber\\
 \var p_j^*            &\simeq& \varepsilon_j^2 + \frac{\varepsilon_j^4}{4} \left\langle
       \frac{\Ex}{\sigma^4} \right\rangle_j.
\label{VarCovarNG}
\end{eqnarray}
It is important that if a large skewness (i.e., asymmetry) of RV errors would be checked
to be negligible, large cross-correlation between $\btheta^*$ and $\bmath p^*$ should not
be expected. The variances of $p_j$ may either increase (for leptokurtic RV errors,
$\Ex>0$) or decrease (for platykurtic RV errors, $\Ex<0$). Note that the
expressions~(\ref{VarCovarNG}) do not require for the shape of distribution of the RV
errors to be known in advance. Provided only the skewness and kurtosis are known, it is
possible to use these expressions in practice. For instance, if the kurtosis of RV errors
is constant, the variance of the corresponding jitter estimation should increase by the
factor $(1+\Ex/2)$. The expressions~(\ref{VarCovarNG}) were checked by means of Monte
Carlo simulations, assuming different simple non-Gaussian distributions for simulated RV
errors (e.g. uniform one). The predictions of analytic formulae~(\ref{VarCovarNG}) for the
jitter estimation variance were found to be in an excellent agreement with results of
numerical simulations, at least for $N$ as big as a few hundred.

Of course, we should satisfy certain conditions of regularity for the theoretical results
described above to be appliable. The rigorous formulation of these conditions is given in
the Appendix~1 by \citet{Gourieroux84}. These incorporate:
\begin{enumerate}
\item Certain requirements of boundedness and integrability for the distribution of the RV
errors (roughly speaking, too heavy tails are not allowed).

\item Conditions of smoothness and boundedness for the equations of the model (RV) curve
(and also for the model of variances, but our square-additive and multiplicative models of
uncertainties are very simple and certainly satisfy them).

\item Requirement that (roughly) any \emph{single} observation should not strongly affect
the final estimations. This put certain condition of `naturality' on the sequence of
observational timings and statistical weights (and also on the RV model).
\end{enumerate}
In fact, these conditions are not qualitatively new. The first one originates from the
requirement of the central limit teorem. The second one originates from the regularity
conditions from the maximum likelihood estimations theory. The third one originates from
both fields.

However, I could not find enough information in the literature about skewness and kurtosis
of RV errors in current planet search surveys. By this reason, I could apply only
formulae~(\ref{Var}), valid for Gaussian RV errors, when calculating the uncertainties of
estimations in Section~\ref{sec_apps}. It is important to note that still the degree of
possible non-Gaussianity of the RV errors in planet searches is not clearly estimated. In
a recent study of the Keck RV survey, \citet{Cumming08} did not reveal clearly any strong
non-Gaussianity.

Non-Gaussian errors may lead to extra $\mathcal O(1/N)$ biasing of $\btheta^*$ and $\bmath
p^*$. This extra bias can be calculated till the first order using the same approach based
on the Taylor expansion of $\ln\mathcal L(\bxi)$ as in Section~\ref{sec_bias}. A non-zero
skewness of RV errors leads to an extra bias $\sim \As/N$ in estimations of $\btheta$
(including estimations of planetary parameters). Considering that very large skewness of
RV errors is unlikely, the `Gaussian' part of the bias in $\btheta$ should dominate in
practice. Therefore, the bias in $\btheta$ inspired by non-Gaussain errors should not be a
practical trouble. A kurtosis excess of RV errors adds some extra bias $\sim \Ex/N$ in the
estimations of jitter. However, this effect is expected to be negligible even for the
kurtosis excess as large as $\Ex= 1 - 3$. For instance, this bias vanishes when the
kurtosis is constant. In any case, the first-order bias in parameters $\btheta$ can be
removed by the Quenouille's algorithm (see Section~\ref{sec_bias}).

\section{Numerical calculation}
\label{sec_numcalc}
It is necessary to propose numerical algorithms performing maximization of the
function~(\ref{log-likelihood-mod}). For the sake of simplicity, let us put $r=1$,
$p=\sigma_\star^2$. The extension to the case $r>1$ will be straightforward and easy. Let
us consider the function $g = \const - 2 \ln \tilde{\mathcal L}$ to be minimized by
$\btheta$ and $p$:
\begin{equation}
g(\btheta,p) = \sum_{i=1}^{N} \left[ \ln\left(1+\frac{p}{\sigma_{{\rm meas},i}^2}\right)
  + \frac{(v_i-\mu(t_i,\btheta))^2}{\gamma(\sigma_{{\rm meas},i}^2+p)} \right].
\label{func_g}
\end{equation}
This function is formally defined for $p > p_0 = - \min \sigma_{{\rm meas},i}^2$. Note
that negative values of $p$ are not senseless. They indicate that the instrumental
uncertainties specified are in fact overestimated.

It seems better to minimize $g(\btheta,p)$ in two steps. In the first step, we will obtain
a second-level target function $h(\btheta) = \min_p g(\btheta,p) = g(\btheta,
p^*(\btheta))$, where $p^*(\btheta)$ denotes the value of $p$ for which this minimum is
achieved. It is evident that in non-degenerated situations $\lim_{p\to+\infty}
g(\btheta,p) = +\infty$ and $\lim_{p\to p_0} g(\btheta,p) = +\infty$. Therefore, for any
fixed $\btheta$ at least one minimum by $p$ exists. The one-dimensional minimization by
$p$ for a fixed $\btheta$ can be precisely and rapidly performed by simple Newtonian-like
algorithms.

A robust situation with only one solution $p^*(\btheta)>p_0$ for a given $\btheta$ usually
takes place. However, sometimes we may deal with the following ill conditioned case.
Suppose that for some $\btheta=\btheta_0$ the residual corresponding to $\sigma_{{\rm
meas}, i} = -p_0$ vanishes. Then the function~(\ref{func_g}) has no global minimum. In
this case, $\lim_{p\to p_0} g(\btheta_0,p) = -\infty$ and the respective solution
$\btheta=\btheta_0$ and $p=p_0$ is not physically sensible. For well-conditioned cases,
real minimization algorithms converge to good solutions which are far from these
singularities. The situations when a numerical algorithm falls in the singularity are very
seldom in practice and appear when the RV uncertainties span a wide range and/or the RV
models are overloaded (contain too many free parameters) and/or they are close to being
degenerate. These cases represent a numerical problem and should be identified during the
minimization. A simple test $1+p/\sigma_{{\rm meas},i}^2<0.01$ is sufficient to diagnose
almost all the singular cases.

In the second step, the function $h(\btheta)$ should be minimized. This may be performed
by standard non-linear least squares algorithms like the Levenberg-Marquardt-Gauss one
\citep[\S\S~5.8--5.11]{Bard}. To show this, we need to check that the gradient and the
Hessian matrix both can be calculated in the same way as during the $\chi^2$ minimization.
Firstly, the gradient $h'(\btheta)$ is equal to the partial derivative
$g'_\btheta(\btheta, p^*(\btheta))$. Within the factor $\gamma$, the last partial
derivative is the usual gradient of the $\chi^2$ function (calculated for the jitter
$p^*(\btheta)$). Secondly, we need to check that the Hessian matrix $h''(\btheta)$ can be
calculated using the Gauss' approach. Recall that the full Hessian matrix for the function
$\chi^2(\btheta)$ is given by
\begin{equation}
 2\, \mathbfss Q - 2 \left\langle (v-\mu) \mu''_{\btheta\btheta}/\sigma^2 \right\rangle.
\label{HessChi2}
\end{equation}
The second term in~(\ref{HessChi2}) has magnitude $\mathcal O(\sqrt N)$ and is neglected
in comparison with the first one, which has the magnitude $\mathcal O(N)$. This is the
commonly used Gauss' approach which allows the calculations of second-order derivatives of
$\mu$ to be avoided. It can be shown that the same approximation is valid for the matrix
$\gamma h''(\btheta)$. The exact expression of $\gamma h''(\btheta)$ contains extra terms
having magnitude $\mathcal O(1)$ (i.e., $\mathcal O(N^0)$) only, that is even less than
the second term in~(\ref{HessChi2}).

\section{Testing hypotheses}
\label{sec_test}
Often we need to choose between at least two hypotheses, a base one $\mathcal H$ and an
alternative one $\mathcal K$, based on the RV data. Usually these hypotheses are defined
by some parametric temporal models of the RV curve, $\mu_{\mathcal H}(t,\btheta_{\mathcal
H})$ and $\mu_{\mathcal K}(t,\btheta_{\mathcal K})$. Here vectors $\btheta_{\mathcal H}$
and $\btheta_{\mathcal K}$ contain $d_{\mathcal H}$ and $d_{\mathcal K}$ unknown
parameters. We will assume that $\mathcal H$ is nested in $\mathcal K$, that is
$\btheta_{\mathcal K} = \{\btheta_{\mathcal H}, \btheta\}$ and $\mu_{\mathcal K}(t,
\btheta_{\mathcal K}) = \mu_{\mathcal H}(t, \btheta_{\mathcal H}) + \mu(t,\btheta)$ where
$d=d_{\mathcal K}-d_{\mathcal H}$ quantities $\btheta$ parametrize the model $\mu$ of some
extra RV variability. The parameters $\btheta$ are chosen so that this extra signal
vanishes when $\btheta=0$: $\mu(t,\btheta=0)\equiv 0$. We wish to test whether the
hypothesis $\mathcal H: \btheta=0$ (no signal) is consistent with our RV data or it should
be rejected in favour of the alternative $\mathcal K: \btheta\neq 0$ (signal exists). The
parameters $\btheta_{\mathcal K}$ are supposed to belong to some domain $\Theta_{\mathcal
K}$ in $d_{\mathcal K}$ dimensions. The condition $\btheta=0$ cuts in this domain a
hypersurface $\Theta_{\mathcal H}$ of dimension $d_{\mathcal H}<d_{\mathcal K}$. Thus we
can reformulate our goal as to check, whether the hypothesis $\btheta \in \Theta_{\mathcal
H}$ is consistent with the RV data or it should be rejected in favour of the alternative
$\btheta \in \Theta_{\mathcal K}\setminus \Theta_{\mathcal H}$.

There are many practical tasks which can be embedded in this mathematical framework. For
instance, often we need to test existence of an extra periodic RV variation of a given
frequency or an extra long-term RV trend. The possible extra periodicity may be modelled
as a sinusoidal harmonic, and the possible trend as a linear or quadratic function.

The common tools used to solve such problems are the $\chi^2$ and $F$ tests. The $\chi^2$
test is based of the difference between the $\chi^2$ functions, calculated for the
best-fitting RV models for the hypotheses $\mathcal H$ and $\mathcal K$. To apply the
$\chi^2$ test, we need to know the full RV uncertainties $\sigma_i$. The $F$ test is based
on the ratio of the same $\chi^2$ functions. The $F$ test is more flexible than the
$\chi^2$ one: it can process cases when only the weights $w_i$ are known a priori, and the
RV uncertainties are calculated according to the multiplicative model~(\ref{MultModel}).
The factor $\kappa$ is estimated implicitly in the $F$ test. In our case, the RV
uncertainties are given by the square-additive model~(\ref{SquareModel}), and the $F$ test
cannot be applied. The RV jitter $\sigma_\star^2$ has to be estimated explicitly. Doing
so, we can construct the logarithm of the likelihood ratio statistic
\begin{eqnarray}
Z =  \max_{\bmath p, \btheta_{\mathcal K}}\, \ln\mathcal L -
  \max_{\bmath p, \btheta_{\mathcal H}}\, \left. \ln\mathcal L \right|_{\btheta=0}.
\label{loglikrat}
\end{eqnarray}
Here, the maximization of $\ln\mathcal L$ by $\bmath p$ means that the RV jitter are
estimated explicitly, together with the usual parameters of the RV curve, $\btheta$. The
resulting best-fitting values are then used to construct the logarithm of the ratio of the
maximized likelihood functions corresponding to hypotheses $\mathcal H$ and $\mathcal K$.
For the purposes of bias reduction, it is better to use the following modification of the
likelihood ratio:
\begin{equation}
 \tilde Z = \frac{N_{\mathcal K}}{N} \left[ \max_{\bmath p, \btheta_{\mathcal K}} \ln\tilde{\mathcal L}_{\mathcal K} -
  \max_{\bmath p, \btheta_{\mathcal H}} \left.\ln\tilde{\mathcal L}_{\mathcal H}\right|_{\btheta=0} \right] +
  \frac{N_{\mathcal K}}{2} \ln \frac{N_{\mathcal H}}{N_{\mathcal K}},
 \label{loglikratmod}
\end{equation}
where $N_{\mathcal H} = N-d_{\mathcal H}$ and $N_{\mathcal K} = N - d_{\mathcal K}$. The
modified likelihood functions $\ln \tilde{\mathcal L}$ are different for hypotheses
$\mathcal H$ and $\mathcal K$, because they contain different correctors $\gamma_{\mathcal
H} = N_{\mathcal H}/N$ and $\gamma_{\mathcal K} = N_{\mathcal K}/N$. Note that if the
multiplicative model were assumed for $\sigma_i$, the function (\ref{loglikratmod}) would
coincide with the statistic $z_3$ from the paper \citep{Baluev08a}. The square-additive
model of $\sigma_i$ generates another form of $\tilde Z$, which is preferred for testing
statistical hypotheses in RV planet search surveys. Note that
definitions~(\ref{loglikrat}) and~(\ref{loglikratmod}) do not require strict linearity of
the models.

A large value of the statistic $\tilde Z$ indicates that the base hypothesis may be wrong
and the specified alternative model is more realistic. However, random RV errors may also
produce similar values of $\tilde Z$. To compute statistical significance of the observed
value of $\tilde Z$ we should know the distribution of $\tilde Z$ under the base
hypothesis $\mathcal H$. Of course, there is a little hope that this distribution can be
calculated exactly. Nevertheless, many asymptotic ($N\to\infty$) results are known for the
likelihood ratio statistic. In particular, the distribution of the quantity $2Z$ (as well
as $2\tilde Z$) converges to the $\chi^2$ distribution with $d$ degrees of freedom, if
certain regularity conditions are satisfied \citep[e.g.][]{Protassov02,Sen79}. Some of
these regularity conditions are technical and are satisfied in the majority of
applications. However, other conditions may not be satisfied in many practical cases, and
therefore they deserve to be checked before applying the asymptotic $\chi^2$ distribution
to $\tilde Z$. It is worth noting that:
\begin{enumerate}
\item The spaces of parameters should be nested, $\Theta_{\mathcal H} \subset
\Theta_{\mathcal K}$. This requirement is already built in our formulation of the
hypothesis testing problem.
\item The subspace $\Theta_{\mathcal H}$ should lie in the interior of $\Theta_{\mathcal
K}$. It should not lie on the boundary of $\Theta_{\mathcal K}$. Otherwise, the asymptotic
distribution of the likelihood ratio statistic is not the $\chi^2$ distribution with $d$
degrees of freedom \citep{Protassov02}. See the paper by \citet{Self87} for a general
algorithm of constructing the asymptotic distribution of $Z$ (or $\tilde Z$) in this
non-standard case. Typically, when $\Theta_{\mathcal H}$ lie on the boundary of
$\Theta_{\mathcal K}$, the asymptotic distribution of the likelihood ratio statistic
appears to be some mixture of $\chi^2$ distributions with different numbers of degrees of
freedom, but more complicated cases are also possible.
\item Equations of the RV models, $\mu_{\mathcal H}(t,\btheta_{\mathcal H})$ and
$\mu(t,\btheta)$, should satisfy certain conditions of smoothness and boundedness.
\end{enumerate}
Note that the same (or similar) regularity conditions are equally required to hold true
when using the $F$ test and the $\chi^2$ test as well. These conditions may not to hold
true for a given parametrization but simultaneously may be satisfied for some other one.
The likelihood ratio statistic and its distribution are invariant with respect to a
re-parametrization. Hence, it is sufficient for the regularity conditions to hold true for
only one parametrization.

Suppose that all the necessary conditions are satisfied, and the distribution of $2\tilde
Z$ indeed converges to the $\chi^2$ one. This convergence is not uniform. Larger values of
$\tilde Z$ correspond to larger displacements in the parameter space. These increase
non-linear effects and require larger $N$. I have followed the convergence of $\tilde Z$
to the $\chi^2$ distribution for the square-additive model~(\ref{SquareModel}) by means of
Monte-Carlo simulations for various structures of time series and simple RV models. The
simulations yielded the following empirical convergence condition:
\begin{eqnarray}
N \gtrsim \rho \tilde Z s \quad {\rm (valid~for~}N>30, \FAP>10^{-5}, s<1}{\rm ),
\label{convcond}
\end{eqnarray}
where $s=\ln(\max\sigma/\min\sigma)$, $\rho=2.5-3$ when testing RV model `constant
velocity' vs. `constant + linear trend' ($d_{\mathcal H}=d=1$), and $\rho=5-6$ for RV
models `constant' vs. `constant + sinusoidal harmonic with a fixed frequency'
($d_{\mathcal H}=1, d=2$). The condition~(\ref{convcond}) is not stringent. The extra RV
jitter softens differences between $\sigma_i$, so that $s<1$ usually. Note that we are
practically interested in the values $\tilde Z = 3-10$ producing $\FAP = 10^{-3}-0.1$ (for
small $d$).

\section{Likelihood periodograms}
\label{sec_periods}
The \citet{Lomb76}--\citet{Scargle82} periodogram and its normalizations require fixed
weights of observations; hence, they assume multiplicative model for RV uncertainties. For
planet searches, it is preferred to use a periodogram with built-in estimation of the
square-additive RV jitter. Such periodogram may be constructed from the modified
likelihood ratio statistic $\tilde Z$ in the same way as it is described in
\citep{Baluev08a} for the $\chi^2$ statistic. The resulting periodogram $\tilde Z(f)$ will
be a function of the frequency $f$ of a trial periodic RV signal (modelled by a sinusoidal
harmonic or by some other periodic function), so that every single value $\tilde Z(f)$
represents the statistic~(\ref{loglikratmod}).

To assess statistical significance of candidate periodicities, we should know the
distributions of the likelihood ratio periodograms. In practice, orbital period of a
possible planet is not known a priori and a wide frequency range is scanned in order to
find the maximum periodogram peak. By analogy with single-value distributions, we can
expect that distributions of maximum values of $\tilde Z(f)$ converges (for $N\to \infty$)
to the distribution of the maximum of the least squares periodogram $z(f)$ from
\citep{Baluev08a}. Here we should be careful, because the convergence of distributions of
maximum values of periodograms is not necessary achieved in practice
\citep{SchwCzerny98a}. Recall that the convergence condition for the periodogram $z_3(f)$
from the work~\citep{Baluev08a} is $z_3\ll N$. This periodogram is a special case of the
likelihood ratio periodogram for the multiplicative model of RV uncertainties. Therefore,
we can expect that our likelihood ratio periodogram with a built-in jitter estimation
requires a convergence condition similar to~(\ref{convcond}). Results of Monte-Carlo
simulations are consistent with this assumption if $\rho=5-6$ or even less (depending on
the aliasing and on the desirable precision of false alarm probability). Unfortunately,
this problem appears too complicated to be discussed here in more detail and probably
requires a separate investigation in future.

\section{Application to real data}
\label{sec_apps}
The mathematical tools described above may be applied to real RV data from planet search
surveys. For this purpose, I have selected several planetary systems with well-determined
orbital configurations and large series of observations from different observatories. RV
data were taken from the works by \citet{Butler06} for the stars 51~Peg, 70~Vir, 14~Her,
HD83443, 54~Psc, $\mu$~Ara; by \citet{Naef04} for 51~Peg, 70~Vir, 14~Her, 55~Cnc; by
\citet{Mayor04} for HD83443; by \citet{Wittenmyer07a} for 14~Her; by \citet{Pepe07} for
$\mu$~Ara; by \citet{Lovis06} for HD69830; by \citet{Wittenmyer07b} for 54~Psc; by
\citet{McArthur04} and by \citet{Fisher08} for 55~Cnc.

\begin{table}
\caption{RV jitter for several stars with planetary systems.}
\label{tab_jitter}
\begin{tabular}{@{}l@{}c@{}c@{}c@{}c@{}c@{}c@{}}
\hline
System     & Instr.$^4$ & $N$ & nomin. ${\sigma_\star}^1$ & $A$~[m/s]   & $\tau$~[date] & resid. ${\sigma_\star}^1$ \\
\hline
51~Peg$^3$ & ELD        &$153$& $9.93(88)$          & $11.2(1.2)$ &Jun~$28.5(9.3)$& $6.56(83)$          \\
b          & LCK        &$256$& $0.53(2.37)$        &             &               & $0.45(2.81)$        \\
\hline
70~Vir     & ELD        & $35$& $-3.0(1.6)^2$       &             &               & $-2.8(1.8)^2$       \\
b          & LCK        & $74$& $4.22(80)$          & $5.1(1.6)$  & May~$16(16)$  & $3.58(79)$          \\
\hline
14~Her$^3$ & ELD        &$119$& $7.32(96)$          & $6.4(1.6)$  & Jan~$6(12)$   & $6.45(96)$          \\
           & KCK        & $49$& $1.25(44)$          &             &               & $1.23(45)$          \\
b          & HJS        & $35$& $-4.46(84)^2$       &             &               & $-4.44(86)^2$       \\
\hline
HD69830    & HRP        & $74$& $0.22(26)$          &             &               &                     \\
b, c, d    &            &     &                     &             &               &                     \\
\hline
HD83443$^3$& COR        &$257$& $6.83(49)$          & $5.8(1.4)$  &Oct~$12.5(9.5)$& $6.40(48)$          \\
           & AAT        & $23$& $5.9(1.7)$          &             &               & $6.0(1.7)$          \\
b          & KCK        & $28$& $2.74(57)$          &             &               & $2.70(58)$          \\
\hline
54~Psc     & LCK        &$121$& $5.78(52)$          &             &               & $5.87(54)$          \\
           & KCK        & $42$& $3.86(52)$          & $4.73(90)$  & May~$27(13)$  & $2.30(40)$          \\
b          & HET        & $83$& $10.1(1.3)$         & $11.7(3.1)$ & Dec~$18(27)$  & $7.8(1.0)$          \\
\hline
$\mu$~Ara  & COR        & $40$& $5.67(95)$          &             &               &                     \\
           & AAT        &$108$& $2.37(25)$          &             &               &                     \\
b, c, d, e & HRP        & $86$& $1.52(15)$          &             &               &                     \\
\hline
55~Cnc$^5$ & ELD        & $48$& $14.2(1.9)$         & $8.3(3.5)$  & Oct~$21(34)$  & $13.3(1.8)$         \\
           & LCK        &$250$& $5.19(33)$          &             &               & $5.19(33)$          \\
b, c, d,   & KCK        & $70$& $4.33(36)$          &             &               & $4.33(37)$          \\
e, f       & HET        &$119$& $6.95(86)$          & $8.2(1.8)$  & Mar~$14(15)$  & $5.22(74)$          \\
\hline
\end{tabular}\\
The uncertainties of the estimations are given in the parentheses in the units of last
digits. For instance, $7.32(96)$ means $7.32\pm 0.96$, and $3.0(1.6)$ means $3.0\pm 1.6$.
The quantities $A$ and $\tau$ are the semi-amplitudes and the epochs of the maximum RV of
the best-fitting sinusoidal annual drifts.\\
$^1$ Uncertainties of jitter estimations are calculated in the linear Gaussian
approximation as $\delta = \varepsilon/ (2\sigma_\star)$, where $\varepsilon$ is the
asymptotic uncertainty of the estimation of $\sigma_\star^2$. When $\sigma_\star$ is
comparable with (or less than) $\delta$, its distribution is far from Gaussian. Then it is
necessary to return back to the quantity $\sigma_\star^2$ (not affected by the degeneracy)
and to its uncertainty $\varepsilon = 2\delta\sigma_\star$.\\
$^2$ Negative values of $\sigma_\star$ symbolically reflect that corresponding estimations
of $\sigma_\star^2$ are negative.\\
$^3$ A linear/quadratic trend was included in the RV model.\\
$^4$ ELD = ELODIE, COR = CORALIE, HRP = HARPS, LCK = Lick Observatory, KCK = Keck
Telescope, AAT = Anglo-Australian Telescope, HET = Hobby-Eberly Telescope, HJS = Harlan~J.
Smith Telescope.\\
$^5$ Dynamical fit, taking planetary perturbations into account. The orbits were assumed
coplanar and seen edge-on (other moderate inclinations gave slightly different but similar
results).
\end{table}

For almost every of these stars we have two or more independent RV datasets. The
configurations of orbits in these systems are determined reliably. For each star, we can
write down the model of the RV curve containing a number of free parameters to be
estimated. The joint estimation of the RV curve parameters and of the RV jitter was
performed using maximization of the modified likelihood
function~(\ref{log-likelihood-mod}). The resulting jitter estimations are shown in the
fourth column of Table~\ref{tab_jitter}. These estimations include analytic bias
correction~(\ref{jitter-bias}), though this correction often appeared negligible due to a
large number of observations. The corresponding estimations of planetary parameters are
omitted (perhaps, they are not so interesting here). We can clearly see that RV jitter for
distinct instruments may differ largely, even for one and the same star. Typically, RV
jitter for ELODIE and CORALIE spectrograph's are significantly higher than for other
instruments. The only exception is the star 70~Vir, for which the ELODIE RV jitter is
consistent with zero (two-sigma upper limit is $3.2$~m/s). In contrast, the HARPS jitter
are remarkably smaller, dropping sometimes below $1$~m/s level. Data from Keck, Lick and
Anglo-Australian observatories demonstrate intermediate cases. \citet{Fisher08} estimated
RV jitter for Lick and Keck data for 55~Cnc by $3.0$~m/s and $1.5$~m/s. However, the
corresponding values from Table~\ref{tab_jitter} are $5.19$~m/s and $4.33$~m/s. This
indicates some extra RV instability having standard deviation about $4.1$~m/s. The actual
nature of this extra RV instability is unclear. Surprisingly, the RV jitter for the HJS
data for 14~Her is definitely negative ($\sigma_\star^2 = -(4.5~{\rm m/s})^2$). This
indicates that the RV uncertainties of these measurements, quoted by
\citet{Wittenmyer07a}, are overestimated by about $20\%$. More careful investigation
reveals that the r.m.s. of the RV residuals for the HJS data of 14~Her is $5.85$~m/s. This
value of r.m.s. is quite reliable, because the RV model is well constrained by the data
from two independent teams (from Lick and ELODIE). Simultaneously, the stated instrumental
RV uncertainties of HJS are ranged between $6.4$~m/s and $12.7$~m/s with the geometric
mean of about $7.4$~m/s.

\begin{figure}
\includegraphics[width=84mm]{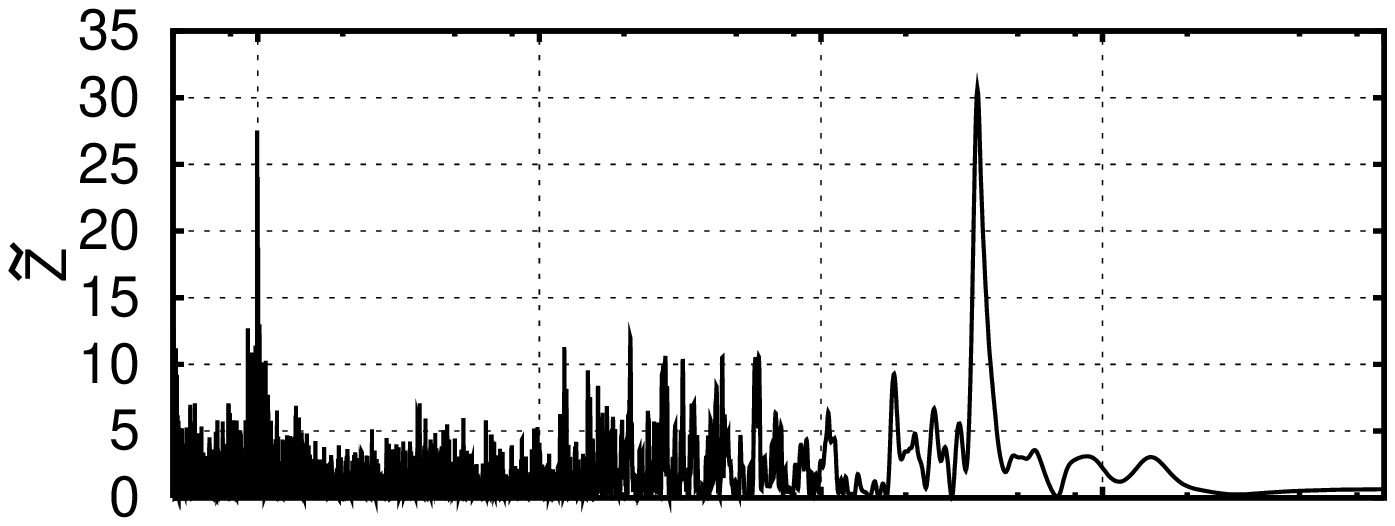}\\
\includegraphics[width=84mm]{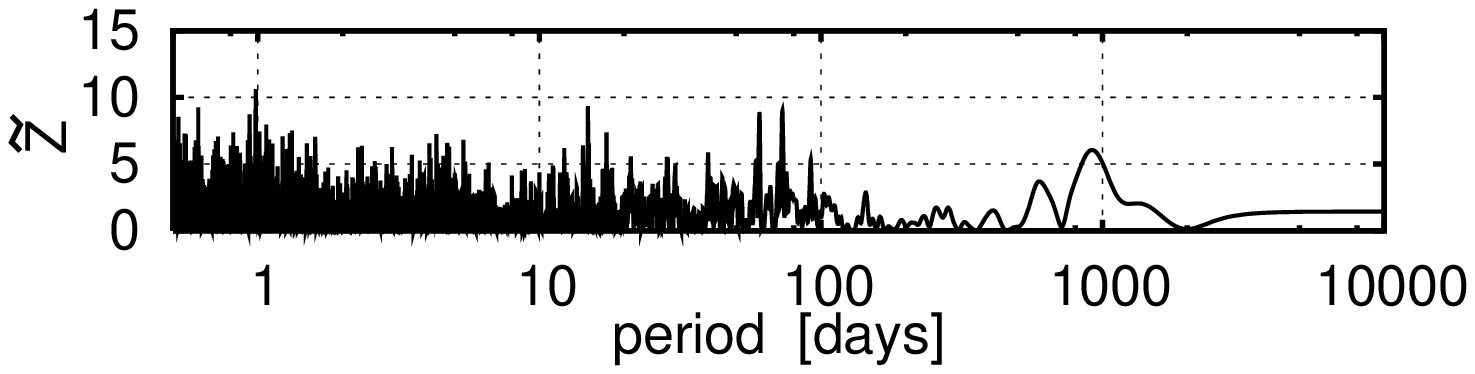}
\caption{Top: likelihood residual periodogram of ELODIE radial velocities of 51 Pegasi.
The RV oscillation induced by the planet 51~Peg b was included in the base model
$\mu_{\mathcal H}$. For the model $\mu$ of the extra signal to be tested, the harmonic
function was adopted (as it is done in the Lomb-Scargle periodogram). The clear peak with
$\tilde Z=30.8$ at the period of $359$~days most likely corresponds to annual instrumental
errors. Three strong peaks near one day ($\tilde Z=27.3$ at $23^h 52^m 6^s$, $\tilde
Z=23.7$ at $24^h 0^m 5^s$, and $\tilde Z=18.4$ at $24^h 4^m 2^s$) may be interpreted as
aliases. Bottom: the periodogram with base RV model `planet b + linear trend + annual
periodicity'. This periodogram is clean.}
\label{fig_51Peg-Annual}
\end{figure}

Large RV jitter for ELODIE and CORALIE data indicate that some systematic errors are not
accounted for in their RV uncertainties. Partially, these systematic errors can be
explained by extra annual RV variations. This can be established by means of including an
extra harmonic term $A \cos(2\pi (t-\tau)/1{\rm yr}) = C \cos(2\pi t/1{\rm yr}) + S
\sin(2\pi t/1{\rm yr})$ in the RV model. Here, the parameters $C,S$ (equivalently
$A,\tau$) were estimated from the RV data together with planetary parameters. Then it can
be checked, whether the maximum value likelihood function is increased statistically
significantly, that is whether the observed value of the corresponding statistic $\tilde
Z$ is statistically significant. As we have $d=2$ (two free parameters of the annual
harmonic), the asymptotic distribution of $\tilde Z$ (and $Z$) is exponential:
$\FAP=\Pr\{\tilde Z>z\}\approx e^{-z}$. Columns 5,6,7 in Table~\ref{tab_jitter} contain
the best-fitting parameters of the annual variations and residual RV jitter, for the
cases when the significance of annual terms was large enough. For the star 51 Pegasi, for
instance, the value of the statistic $\tilde Z$ (almost equal to $Z$ in this case, due to
a very large $N=409$), associated with the annual periodicity, was about $35$. For
comparison, the $3$-sigma significance level is $\tilde Z\approx 5.9$.

We can see that annual RV errors in planet search surveys represent a frequent phenomenon
(although cases free from these systematic errors are also not rare). Possible sources of
such systematic errors may have various nature. For instance, for ELODIE and CORALIE they
may be partially inspired by weak telluric lines which were not completely excluded in old
cross-correlation templates (M.~Mayor, private communication). Keck and Lick data
published before 2006 contain an inaccuracy due to non-relativistic barycentric
correction. This inaccuracy was below $1$~m/s for the majority of stars, but in rare cases
reached $3-5$~m/s (J.T.~Wright, private communication). Sometimes annual periodicities can
produce rather strong peaks on periodograms (Fig.~\ref{fig_51Peg-Annual}). However, a more
dangerous situation takes place when such contaminating RV variations are not seen on the
periodogram clearly. Despite of this fact, they can produce significant distortions of the
best-fitting orbital model of the planetary system. Therefore, often it may be useful to
check how much the best-fitting orbital configuration of a planetary system is changed if
the annual harmonic term is added to the RV model.

\citet{Butler06} report on the detection of an extra RV trend of $-1.64 \pm
0.16$~m/(s$\cdot$yr) in the Lick RV data for 51~Peg. This trend may be induced by an extra
unseen companion on a long-period orbit. However, the ELODIE data alone yield an
estimation of $-0.15 \pm 0.40$~m/(s$\cdot$yr), which is poorly consistent with the Lick
data. Only after addition of an annual harmonic to the model of the ELODIE data the
long-term trend can be confirmed. Significance of this slope, based on the ELODIE data
only, now corresponds to $2.7$-sigma level. Its magnitude of $-0.94 \pm
0.34$~m/(s$\cdot$yr) is now much better consistent with the estimation by
\citet{Butler06}, although some residual difference at the level of less than two sigma
may indicate some extra (probably non-periodic) systematic RV errors, yet to be corrected
or taken into account. The joint estimation, based on the Lick and ELODIE data, yields
$-1.46 \pm 0.16$~m/(s$\cdot$yr).

It is interesting to consider planet candidates having orbital periods consistent with one
year. There are about fifteen such planets. The masses of all of them but one are rather
large (several Jupiter masses), producing large RV semi-amplitudes $\sim 100$~m/s.
However, the planet HD74156 d announced recently on the basis of HET observations
\citep{Bean08} has relatively small mass and induces the RV oscillation of $\sim 10$~m/s
only. Given similar amplitude of the annual periodicity in the HET data for the stars
55~Cnc and 54~Psc, the discovery of HD74156 d looks suspicious. More careful analysis
shows that RV data for HD74156 from ELODIE \citep{Naef04} also show an annual variation of
$\sim 20$~m/s, but in opposite phase. This inspires strong doubts about the existence of
the planet HD74156 d. This planet candidate could quite be a false detection made due to
annual errors in RV data from HET. In any case, its orbital parameters may be strongly
distorted and are unreliable.

The work \citep{Baluev08c} provides a more intricate example, dealing with RV data for the
system around HD37124. It presents a careful application of the methodology described
here. It further demonstrates either the importance of the differences between the
effective RV jitter for ELODIE, CORALIE, and Keck data, or the importance of annual errors
in obtaining suitable orbital solutions.

\section{Conclusions}
\label{sec_conclusions}
In this paper, the problem of poorly constrained radial velocity jitter in planet search
surveys is considered. An algorithm for the RV curve fitting with a built-in accounting
for this jitter is developed. In many cases, this algorithm gives much better accuracies
of estimation of the \emph{full} RV jitter than the usage of only \emph{astrophysical}
jitter estimations based on the empirical correlations with stellar characteristics. This
algorithm is based on the maximum likelihood principle and includes a series of bias
corrections, either analytic or numerical. An extension of the Lomb-Scargle periodogram
with a built-in jitter estimation is proposed. An effect of non-Gaussian RV errors is
discussed and is shown to be tolerable for large time series. It was shown that numerical
computations based on this method may be implemented by means of the usual least squares
Levenberg-Marquardt-Gauss algorithm with slight modifications.

This methodology can be useful for obtaining best-fitting orbital configurations in
extrasolar planetary systems, especially in the case when inhomogeneous RV data (taken
from different observatories) are merged in the analysis. It can also provide better input
data for improving the empirical correlations found in activity models, since the maximum
likelihood approach provide a better accuracy of jitter estimations than the
tradiationally used method of moments (which is based on the difference between observed
scatter of RV residuals and instrumental noise level).

The application of these mathematical tools to several stars with published RV data
revealed that the effective values of the RV jitter may be quite different for different
instruments (even for one and the same star). Probably, the main reason of these
differences is a poor knowledge of instrumental RV errors. In many cases (especially for
ELODIE and CORALIE data), an extra annual harmonic in the RV model leads to significant
improvements in fit quality and somewhat decreases too large effective RV jitter. It is
shown that the planet candidate HD74156 d with orbital period close to one year may be a
false detection made due to annual errors in the HET RV data.

\section*{Acknowledgments}
I would like to thank Drs. V.V.~Orlov and K.V.~Kholshevnikov for critical reading of this
paper, useful suggestions, and linguistic corrections. Dr S.~Ferraz-Mello is thanked for
fruitful suggestions concerning terminology used in the paper. I am grateful to Drs.
M.~Mayor, J.T.~Wright, and A.~Quirrenbach for detailed discussions of possible sources of
annual RV errors in present planet search surveys. I would like to thank especially the
referees, Drs. S.H.~Saar and V.L.~Kashyap, for their extremely deep and helpful analysis
of the manuscript. The work presented in the paper was supported by the Russian Foundation
for Basic Research (Grant 06-02-16795) and by the President Grant NSh-1323.2008.2 for the
state support of leading scientific schools.

\bibliographystyle{mn2e}
\bibliography{jitter}

\appendix

\section{Notation}
\label{sec_not}
Let us introduce the following operation:
\[
 \langle \phi(t) \rangle = \sum_{i=1}^{N} \phi(t_i),
\]
where $t_i$ is an $i^{\rm th}$ observational epoch. Similar summation $\langle a_i
\rangle$ may be defined for a discrete sequence $a_i$. For shortness, the argument $t$ and
the index $i$ are omitted in the text.


All vectors (including gradients) are assumed to be column ones by default. The notation
$\{\bmath x_1,\bmath x_2,\ldots\}$ correspond to a vector constituted by elements of the
vectors $\bmath x_1,\bmath x_2,\ldots$.




If $\bmath x, \bmath y$ are vectors then $\bmath x \otimes \bmath y := \bmath x \bmath
y^T$ is a matrix constituted by the pairwise products $x_i y_j$.


$\var \bmath x := \expect (\bmath x \otimes \bmath x) - \expect \bmath x \otimes \expect
\bmath x$ is the variance-covariance matrix of the random vector $\bmath x$, and
$\cov(\bmath x,\bmath y) := \expect (\bmath x \otimes \bmath y) - \expect \bmath x
\otimes \expect \bmath y$ is the cross-covariance matrix of $\bmath x$ and $\bmath y$.



\bsp

\label{lastpage}

\end{document}